\PassOptionsToPackage{hypertexnames=false}{hyperref}
\documentclass[pdflatex,sn-mathphys-num]{sn-jnl}

\usepackage[utf8]{inputenc}
\usepackage{graphicx}
\usepackage{multirow}
\usepackage{amsmath,amssymb,amsfonts}
\usepackage{amsthm}
\usepackage{mathrsfs}
\usepackage[title]{appendix}
\usepackage{xcolor}
\usepackage{textcomp}
\usepackage{manyfoot}
\usepackage{booktabs}
\usepackage{algorithm}
\usepackage{algorithmicx}
\usepackage{algpseudocode}
\usepackage{listings}
\usepackage{ulem}
\usepackage{bibunits}

\newcommand{\eq}[1]{Eq.~\eqref{eq:#1}}

\newcommand{\fig}[1]{Fig.~\ref{fig:#1}}

\newcommand{\ket}[1]{{\mid} {#1}{\rangle}}

\makeatletter
\let\DualSavedMaketitle\maketitle
\let\DualSavedAtMaketitle\@maketitle
\let\DualSavedTitle\title
\let\DualSavedAuthor\author
\newcommand{\ResetForSupplementaryTitle}{%
  \global\let\maketitle\DualSavedMaketitle
  \global\let\@maketitle\DualSavedAtMaketitle
  \global\let\title\DualSavedTitle
  \global\let\author\DualSavedAuthor
  \global\let\artauthors\@empty
  \global\let\auaddress\@empty
  \global\let\corrauthemail\@empty
  \global\let\authemail\@empty
  \global\let\@copycorthanks\@empty
  \global\let\authorsep\@empty
  \global\let\sep\@empty
  \global\aucount=0
  \global\corraucount=0
  \global\punctcount=0
  \global\addcount=0
  \global\emailcnt=0
  \setcounter{affn}{0}%
  \setcounter{footnote}{0}%
  \global\let\@abstract\@empty
  \global\let\@keywords\@empty
  \global\let\@artnote\@empty
  \global\let\@miscnote\@empty
  \global\@auemailfalse
  \global\@corauemailfalse
}
\makeatother

\begin{document}

\begin{bibunit}[sn-mathphys-num]
\title[Article Title]{Magic-wavelength matter-wave interferometry with optical clock states}

\author[1]{\fnm{Jianing} \sur{Li}}

\author[1,2,3]{\fnm{Swarup} \sur{Das}}

\author[3]{\fnm{Xinyuan} \sur{Ma}}

\author[4]{\fnm{Thomas} \sur{Zanon-Willette}}

\author[1,5,6]{\fnm{Shau-Yu} \sur{Lan}}

\author*[1,2,3]{\fnm{Chang Chi} \sur{Kwong}}\email{changchikwong@ntu.edu.sg}

\author[1,2,3]{\fnm{David} \sur{Wilkowski}}

\affil[1]{\orgdiv{Nanyang Quantum Hub, School of Physical and Mathematical Sciences}, \orgname{Nanyang Technological University}, \orgaddress{\street{21 Nanyang Link}, \city{Singapore}, \postcode{637371}, \country{Singapore}}}

\affil[2]{\orgdiv{MajuLab, International Joint Research Unit IRL 3654}, \orgdiv{CNRS, Universit\'e C\^ote d'Azur, Sorbonne Universit\'e, National University of Singapore, Nanyang Technological University} \country{Singapore}}

\affil[3]{\orgdiv{Center for Quantum Technologies}, \orgname{Nanyang Technological University Singapore}, \orgaddress{\street{50 Nanyang Avenue}, \city{Singapore}, \postcode{639798}, \country{Singapore}}}

\affil[4]{\orgname{Sorbonne Universit\'e CNRS, MONARIS, UMR 8233}, \orgaddress{\postcode{F-75005}, \state{Paris}, \country{France}}}

\affil[5]{\orgname{Department of Physics, National Taiwan University}, \orgaddress{\state{Taipei}, \country{Taiwan}}}

\affil[6]{\orgname{Institute for Atomic and Molecular Sciences, Academia Sinica}, \orgaddress{\state{Taipei}, \country{Taiwan}}}

\abstract{Optical clocks and atom interferometers provide complementary ways to measure time, motion and gravity. Combining these capabilities requires matter-wave beam splitters that manipulate different clock states in the same way, so that optical internal energy becomes a controlled degree of freedom rather than a source of systematic phase shifts. Here we realized a dual matter-wave interferometer operating simultaneously on the two states of the $^{88}$Sr optical clock transition, $^1S_0$ and $^3P_0$. The interferometer is driven by Bragg pulses at the 813~nm magic wavelength, for which the two clock states experience the same optical coupling strength. This realizes a common matter-wave beam splitter for atoms whose internal energies differ by an optical excitation. With a sensitivity of 30~mrad, our measurement is consistent with a zero differential phase shift between the two clock-state Mach-Zehnder interferometers, translating to an absence of state-dependent acceleration in free fall at the level of $10^{-5}$. We further used the same interferometer to measure state-dependent optical dipole forces and determine a tune-out wavelength of the metastable $^3P_0$ state to be 478.95(8)~nm. These results establish magic-wavelength clock-state interferometry as a platform for differential force sensing, excited-state polarizability metrology and future quantum-clock tests of gravity.
}

\keywords{Atomic Interferometry, Optical atomic clock, Magic wavelength, Weak Equivalence Principle}


\maketitle
\section{Introduction}\label{sec1}
Advances in coherent manipulation of atomic systems have enabled increasingly sensitive measurements for precision metrology and tests of fundamental physics. Prominent examples include optical atomic clocks, based on high-resolution spectroscopy of narrow optical transitions~\cite{RevModPhys.87.637}, and matter-wave interferometers, which use interference between coherently separated atomic wave packets~\cite{RevModPhys.81.1051}. 
Beyond establishing the international standard for time and frequency, atomic clocks have become indispensable tools for testing predictions of special and general relativity~\cite{chou_optical_2010}, searching for possible variations in fundamental constants~\cite{PhysRevLett.90.150801,doi:10.1126/science.1154622}, and probing physics beyond the Standard Model~\cite{Wcislo2016}. In parallel, 
matter-wave interferometers have become key devices for precision measurements of Earth's gravitational field either in a gravimeter \cite{Jun_Luo_gravimeter_2013,Muller_gravimeter_2008,A_Peters_gravimeter_2001} or gradiometer \cite{rosi_precision_2014,overstreet_observation_2022} configuration, as well as for rotation sensing {\cite{Riehle1991,Kasevich_rotation_1997,Kasevich_2011}}. In addition, they are used for tests of the equivalence principle in its weak form \cite{schlippert2014quantum,asenbaum_atom-interferometric_2020,zhou_joint_2021}, and for the determination of fundamental constants, such as the universal gravitational  constant \cite{rosi_precision_2014} and the fine-structure constant \cite{parker_measurement_2018,morel_determination_2020}.

Matter-wave interferometry with neutral atoms was developed in the 1990s using both thermal atomic beams~\cite{keith_interferometer_1991} and laser-cooled atomic ensembles~\cite{kasevich_atomic_1991}. In contrast to previous neutron interferometers~\cite{maier-leibnitz_interferometer_1962}, atoms are polarizable particles possessing internal quantum states allowing for coherent manipulation of both their internal (e.g. clocks) and external (e.g. Bragg spectroscopy) degrees of freedom through optical light pulses~\cite{Borde1989,Borde1991}. As a matter of fact, in matter-wave interferometry, internal states have been essentially used to facilitate manipulations and detection by stimulated-Raman scattering \cite{kasevich_atomic_1991} or by single-photon excitation between two long-lived states \cite{rudolph2020large}. However, it is possible to further leverage the internal structure of an atom to perform atomic clock interferometry~\cite{Zych2011,loriani_interference_2019}, where the internal states act as an additional resource for sensing. 

In this article, we report on a dual Mach–Zehnder interferometer (MZI)  using magic-wavelength Bragg pulses that operate simultaneously on the $^1S_0$ ground state and the $^3P_0$ metastable state of the strontium-88 ($^{88}$Sr) clock transition~\cite{takamoto_optical_2005}. Magic wavelength Bragg pulses ensure that the light-matter interaction is state independent~\cite{ma2025magic}, whereby in the non-relativistic limit \cite{loriani_interference_2019},  bias phase shifts induced by the optical field are rejected in the common-mode signal~\cite{Akatsuka_2017}. Thus, as expected, we did not observe a differential phase shift down to a sensitivity of $30\,$~mrad when atoms evolve in free fall. 

Using the energy-mass equivalence, we interpret the above result as a test of the weak equivalence principle with a mass difference of $\Delta m=\hbar\omega/c^{2}=1.9\times 10^{-9}\,$u, solely due to the energy difference of the $^1S_0\rightarrow\,^3P_0$ clock transition at a wavelength of $\lambda=2\pi c/\omega=698\,$nm. $\omega$ and $c$ are the angular frequency of the clock transition and the speed of light in vacuum, respectively. We report an Eötvös ratio of $\eta=0.6(12)\times10^{-5}$. A similar weak equivalence principle experiment was previously reported in a non-magic configuration with an internal-energy splitting about five orders of magnitude smaller using the hyperfine structure of rubidium atoms \cite{rosi_quantum_2017}. 

Finally, we applied the dual MZI to probe the state-dependent phase accumulation due to an off-resonant laser that is turned on during the interferometer free-evolution time. We found a tune-out wavelength at $478.95(8)$~nm corresponding to a vanishing polarizability of the $^3P_0$ excited state \cite{safronova2015extracting}. Several tune-out frequencies have been reported using either interferometers with thermal \cite{holmgren2012measurement,decamps2020measurement} and ultracold atoms \cite{copenhaver2019measurement}, or lattice excitation \cite{heinz2020state,ratkata2021measurement,wen2021experimental,PhysRevA.105.L030802,de2025dissipationless, Hohn_2026}. Among all these measurements, only Ref.~\cite{Hohn_2026} has measured the tune-out wavelength for an excited state.

\section{Results}\label{sec2}

\begin{figure}[t]
\includegraphics[width=1\textwidth]{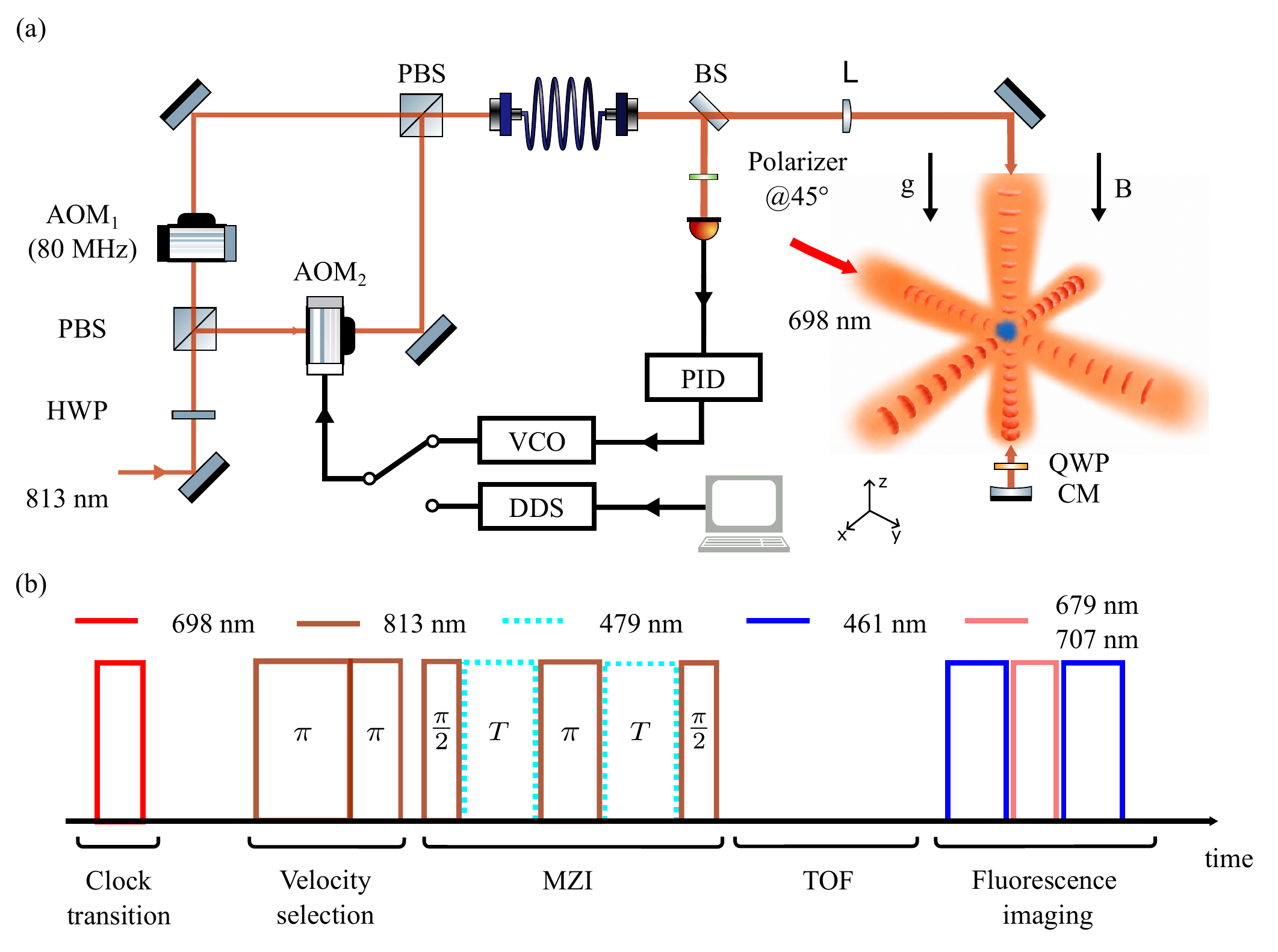}
\caption{\textbf{Schematic of the optical setup, vertical beams phase control, and temporal sequence}. (a) The vertical lattice 813 nm laser beam is initially split into two cross-polarized beams using a half-wave plate (HWP) and a polarizing beam splitter (PBS). The beams are independently modulated by acousto-optic modulators AOM$_1$ (fixed frequency at 80 MHz) and AOM$_2$. They are then recombined, fiber-coupled, and focused onto the atomic cloud with a $750\,$mm lens (L) to realize a waist of $75\,\mu$m at the atoms (blue circle). The beam polarizations are swapped by a quarter-wave plate (QWP) before being refocused back to the atomic cloud thanks to a curved mirror (CM). The AOM$_2$ is driven by a voltage-controlled oscillator (VCO) or a direct digital synthesiser (DDS) for optical lattice and Bragg diffraction operations, respectively. For lattice operation, a fraction of the total power is reflected by a beam sampler and interfered after passing through a $45^\circ$ polarizer for phase lock. For Bragg diffraction, the frequency chirp is digitally controlled. The two horizontal lattices are each composed of a single retro-reflected beam. Their optical paths are not drawn for clarity purposes. The clock laser at 698 nm is overlapped with one of the horizontal lattice beams. (b) Schematic temporal sequence showing the laser pulses and the key events of the experiment.}\label{fig:fig1} 
\end{figure} 

Our experiment begins with a bosonic  $^{88}$Sr ultracold gas prepared using a standard two-stage magneto-optical trap operating first on the $^{1}S_{0}\!\to\!^{1}P_{1} $ transition at 461 nm (linewidth 32 MHz), followed by the $^{1}S_{0}\!\to\!^{3}P_{1}$ transition at 689 nm (linewidth 7.5 kHz) \cite{li_bi-color_2022}. The cold gas is then loaded into a three-dimensional (3D) far-off-resonant optical lattice tuned to the $\lambda_m= 813.4276(2)\,$nm magic wavelength for the $^1S_0 \to\,^3P_0$ clock transition \cite{Katori_magic_PhysRevA.81.023402}. Here, we trapped $N=1.0(1)\times10^6$ atoms at a temperature of $T_0=0.32(2)\,\mu$K, measured via a time-of-flight experiment after adiabatically switching off the lattices \cite{kastberg1995adiabatic}. 

A 698 nm laser resonant with the clock transition is superimposed with one of the horizontal lattices. The 698 nm beam is turned on for 50 ms (Rabi frequency  $\sim100$ Hz), leading to a mixed state with almost equal populations in the $^1S_0$ and $^3P_0$ state. The $^{88}$Sr doubly-forbidden clock transition is enabled via magnetically induced coupling, using a 6 mT bias magnetic field \cite{taichenachev_magnetic_2006,Baillard:07}.

The vertical lattice is formed by two beams whose frequencies are independently controllable (see \fig{fig1}a). If the frequency difference $\Delta\omega$ is set to zero and the beams are phase locked (see  Methods and Supplementary Information), the standing wave is static, and is used to hold atoms. Alternatively, the Bragg operations, necessary for velocity selection and MZI, are performed at $\Delta\omega\neq0$ with the horizontal lattices switched off (see \fig{fig1}b) \cite{altin2013precision}. First, we performed a velocity selection sequence composed of one second-order Bragg $\pi$-pulse and one first-order Bragg $\pi$-pulse that couple momentum states between $0.5\hbar k$ and $4.5\hbar k$, and between $4.5\hbar k$ and $6.5 \hbar k$, respectively. $k=2\pi/\lambda_m$ is the angular wavenumber of the lattice beams. The momentum width is reduced to 0.67$\hbar k$, and it comes with a transfer of  $\sim 10\%$ of the atoms. Then we applied the MZI composed of a pulse sequence $\pi/2-\pi-\pi/2$, with a free-evolution time $T$ between the pulses. Finally, the two momentum output ports of the MZI are read out by fluorescence imaging on the $^1S_0 \to\,^1P_1$ broad transition after a $16\,$ms ($20\,$ms) time-of-flight (TOF) to probe the momentum distribution of the atoms in the $^1S_0$ ($^3P_0$) state. Here, after reading out the $^1S_0$ population, atoms in that state are flushed away before atoms in $^3P_0$ are optically pumped to $^1S_0$ for population measurement.

During the MZI sequence, a frequency chirp $\alpha$ is applied so that the frequency difference $\Delta\omega(t)=\Delta\omega_0-\alpha t$ maintains the Bragg resonance condition in the presence of Earth's gravitational acceleration $g$. Here, $t$ is defined relative to the start of the interferometer sequence, and $\Delta\omega_0=30\,\omega_r$ the frequency difference corresponding to the Bragg resonance between the momentum states $\ket{6.5\hbar k}$ and $\ket{8.5\hbar k}$. $\omega_r=\hbar k^2/(2m)=21.5\times10^{3}\,\textrm{rad/s}$ is the recoil angular frequency. The normalized population difference between the MZI-output ports $\ket{8.5\hbar k}$ and $\ket{6.5\hbar k}$ follows,
\begin{equation}
\Delta N_{i}=- V_i\cos{[(k_{\textrm{eff}}a_i+\alpha)T^2]},
\label{eq:fitOutput}
\end{equation}
where $V_i$ is the interferometer visibility of a clock state $i=\{^1S_0,\,^3P_0\}$, $k_\textrm{eff}=2k$ is the angular wavenumber difference between the two Bragg beams, and $a_i$ is the inertial acceleration. 
\begin{figure}[t]
\begin{center}
\includegraphics[width=\textwidth]{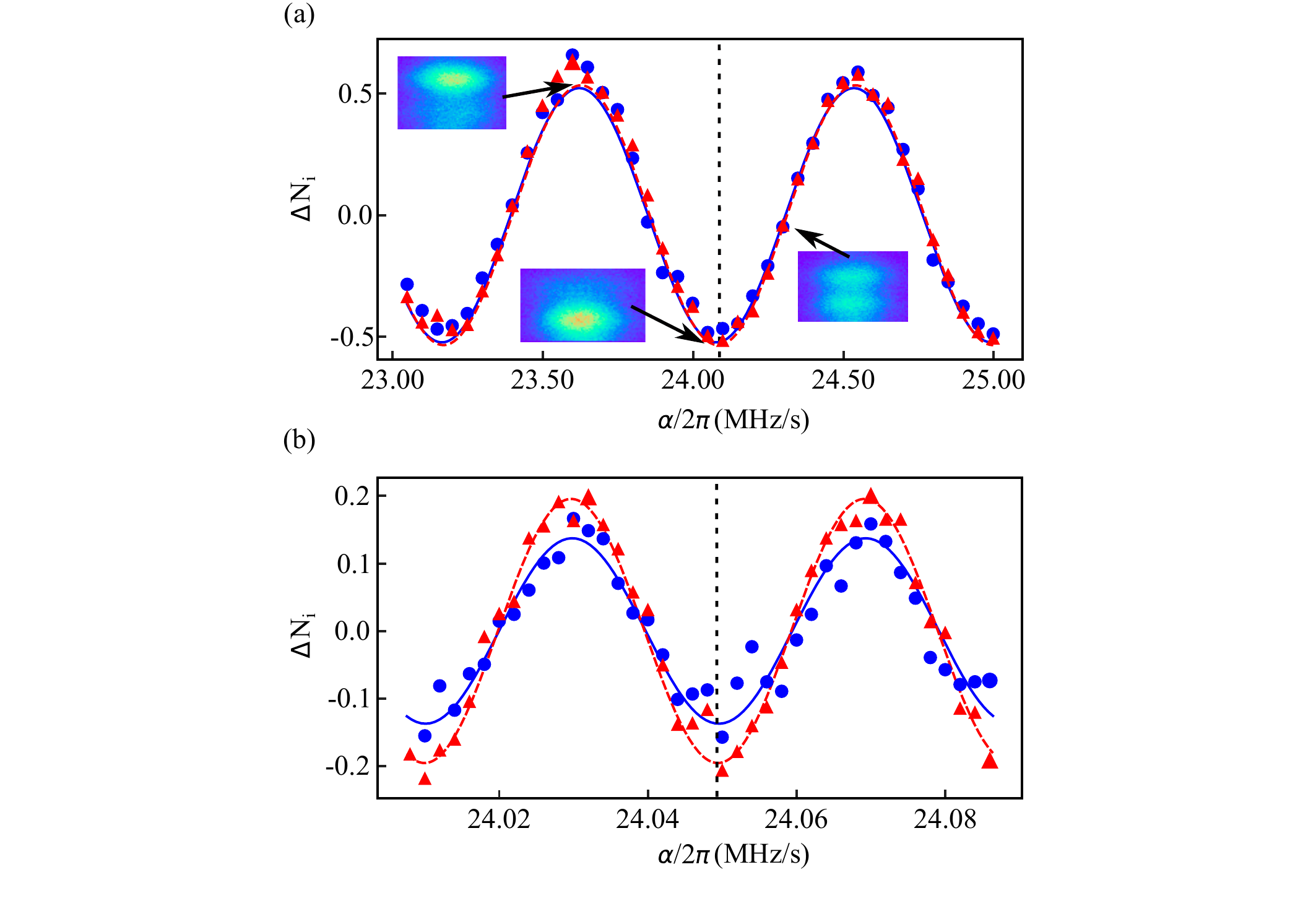}
\caption{ \textbf{Dual interferometric signal.} (a) Normalized population differences $\Delta N_i$ between the output ports for both $^1S_0$ and $^3P_0$ states at a free evolution time of $T=1\,$ms, as a function of frequency chirp of the Bragg laser. The blue dots and red triangles represent the ground and the excited state interferometer population difference, and the solid and dashed curves represent sinusoidal fits using \eq{fitOutput}. Insets: Raw data image after time-of-flight showing the ground state populations. The distribution shows two maxima corresponding to the two output ports of the interferometer. (b) Same as (a) for a free-evolution time of $T=5\,$ms. For (a) and (b), the black short-dashed vertical line corresponds to $\alpha=k_{\textrm{eff}}g$, and each data point is an average of ten experimental runs.} \label{fig:fig2} 
\end{center}
\end{figure} 

In \fig{fig2}a, we show the output population difference for the ground $^1S_0$ (blue dots) and the excited $^3P_0$ levels (red triangles) as a function of the frequency chirp $\alpha$ for a free-evolution time $T=1\,$ms. Similar data for $T=5\,$ms are shown on \fig{fig2}b. This experiment leads to two key results. The first one concerns the local Earth's gravitational acceleration, where $a_i=-g$. We identified a common minimum of the fringes at various values of $T$, where the gravitational acceleration is compensated by the chirp rate, namely, $\alpha_g=k_{\textrm{eff}}g$ (black short-dashed lines in \fig{fig2}). From the ground state data at $T=5\ \text{ms}$, which provides the largest sensitivity, we found Earth's gravitational acceleration of $g=9.78035(12)~ \text{m/s}^2$. The reported uncertainty is statistical in nature, as the systematic errors are yet to be accounted. 

The second result concerns the main consequence of the magic wavelength MZI, which is a null phase difference $\Delta\phi$ between the two clock states, independent of the free-evolution time. We measured $|\overline{\Delta\phi}|=12(12)\,$mrad averaging over measurements within $T\in[1,5]\,$ms (see Supplementary Information for more details). Here, the reported uncertainty corresponds to 1-$\sigma$ statistical error. The null phase difference can also be considered as a test of the weak equivalence principle at a mass difference in the eV/c$^2$ range. The relative differential acceleration measurement is characterized by the Eötvös ratio:
\begin{eqnarray}
\eta=2\left|\frac{a_{\,^1\!S_0}-a_{\,^3\!P_0}}{a_{\,^1\!S_0}+a_{\,^3\!P_0}}\right|.
\label{eq:eotvos}
\end{eqnarray}
Taking the most sensitive realization at $T=5\,$ms, we extracted a null result $\eta=0.6(12)\times10^{-5}$ in agreement with the weak equivalence principle.

\begin{figure}[t]
\begin{center}
\includegraphics[width=\textwidth]{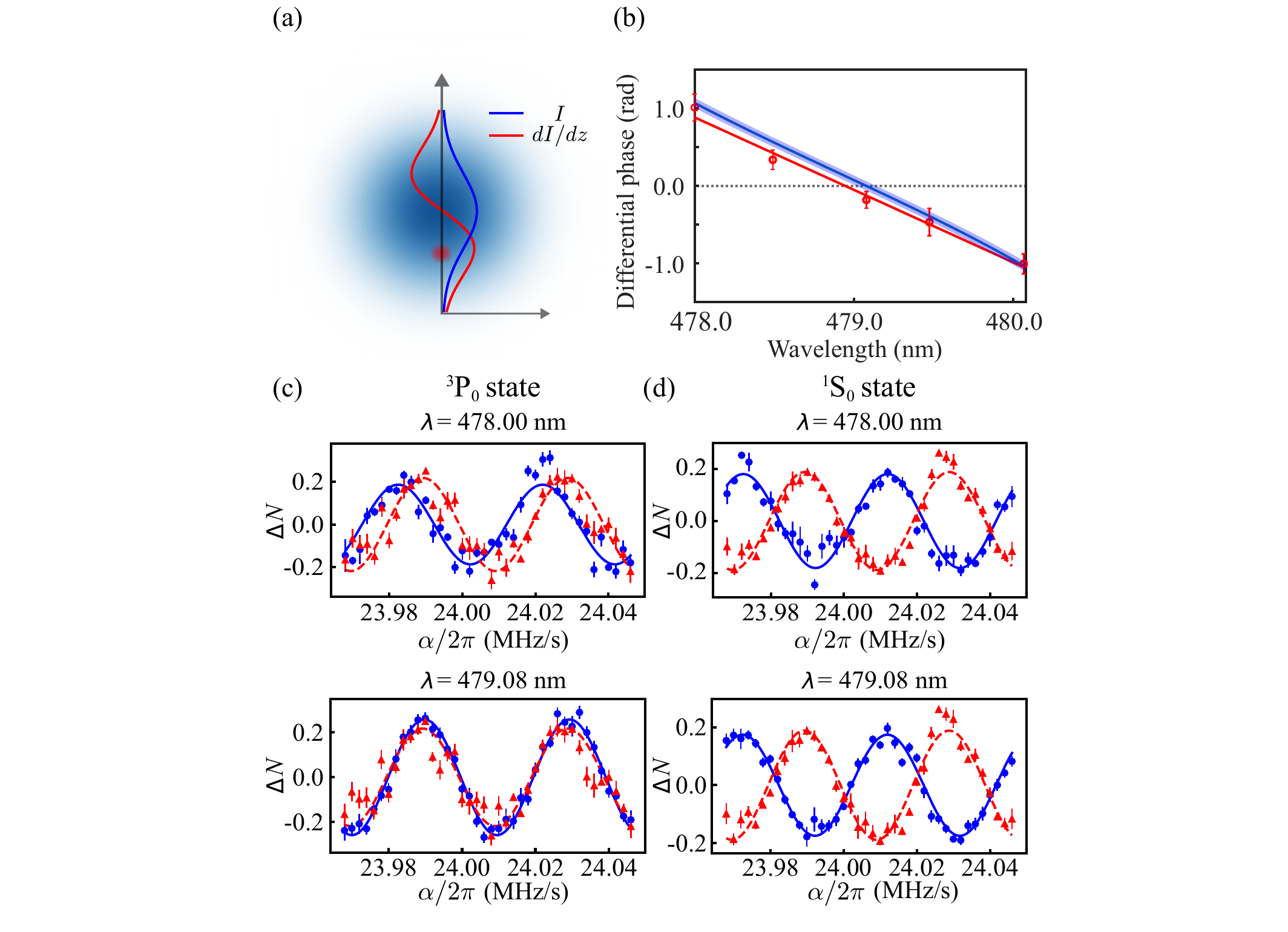}
\caption{\textbf{State-dependent interferometry.} (a) Schematic of the initial position of the atomic cloud (red disk) within the state-dependent off-resonant laser beam (blue blurred disk) at $\sim479\,$nm. The beam is located above the atomic cloud at a distance $\sim w_0/2$ where the force is maximal. The beam waist is $w_0= 620(5)\,\mu$m  and the power of the laser beam is $P=48(2)\,$mW. (b) Relative phase of the excited state MZI performed in the presence of the state-dependent force compared to the phase of the same MZI sequence performed in free fall. The red open circles are the experimental data points, and the red line is a linear fit. The blue curve is the theoretical prediction, and the blue shaded area represents the uncertainty in the theoretical prediction due to uncertainties in the dipole moments of the transitions used to calculate the state polarizability (see Supplementary Information). Uncertainties and error bars correspond to one standard deviation. The horizontal dashed line indicates zero polarizability. (c) Examples of interference fringes involving the $^3P_0$ state. The plot shows the MZI output population difference for an evolution time of $T=5~$ms as a function of frequency chirp of the Bragg laser at wavelengths of $478.0~$nm (top graph) and $479.08~$nm (bottom graph). The blue circles and red triangles are the experimental population differences when the additional optical gradient field is turned on and off, respectively. The plain curves are the fits using \eq{fitOutput}, where $a_i=-g+\Delta\varphi_i/(k_{\textrm{eff}}T^2)$ with $\Delta\varphi_i$ the differential phase plotted in (b) for the excited state. (d) same as (c) but for the $^1S_0$ state. } \label{fig:fig3} 
\end{center}
\end{figure} 

The dual MZI, presented above, can be viewed as a differential accelerometer that probes the local state-dependent inertial forces. We use this property to explore the light interaction of an off-centered $479\,$nm laser beam as shown in \fig{fig3}a. Since the two clock states have different parity and a large energy separation, the differential polarizability $\Delta\alpha$ can also be large, leading to a measurable force in the sensitivity range of our device. This is indeed the case for strontium atoms at $479\,$nm, where the polarizabilities are $\alpha_{\,^1\!S_0}\simeq 2500\,$a.u. and $\alpha_{\,^3\!P_0}\simeq 0\,$a.u. for the $^1S_0$ and $^3P_0$ states, respectively \cite{Safranova_PhysRevA.87.012509}. We placed a beam with waist $w_0=620(5)\,\mu$m and power $P = 48(2)\,$mW above the atomic gas, ideally located at $\sim w_0/2$ away from the beam center along the vertical axis to get a maximal state-dependent force, as depicted in \fig{fig3}a (see Supplementary Information for more details). In \fig{fig3}b, we present several measurements of the MZI phase shift for the excited state as a function of the laser frequency. Each data point in \fig{fig3}b is extracted from fits of the Mach-Zehnder interferometry fringes, with characteristic examples shown in \fig{fig3}c. The phase shifts observed in \fig{fig3}b are in quantitative agreement with the polarizability values derived following ref.~\cite{Safranova_PhysRevA.87.012509}. The tune-out wavelength is determined by a linear fit and found to be $478.95(8)$\,nm, in agreement with the predicted value of 479.08(1) nm \cite{safronova2015extracting}. As shown in  \fig{fig3}d, the MZI phase shift for the $^1S_0$ state is large and almost constant, as expected since $d\alpha_{\,^1\!S_0}/d\lambda(479\text{ nm})\simeq 130\text{ a.u./nm}$, {\it i.e,} about eight times smaller than for the $^3P_0$ state.

\section{Conclusion and Discussion}\label{sec12}

In summary, we reported on a state-differential dual Mach-Zehnder interferometer (MZI) operating simultaneously on the $^{88}$Sr clock states. To avoid state-dependency at the matter-wave manipulation level, we operated the Bragg pulses of the MZI at the $813\,$nm magic wavelength. Hence, in a free-fall configuration, we observed no differential phase shift. This result can be interpreted as a test of the weak equivalence principle, where the mass difference is solely due to the energy difference between the two states. We found an E\"otv\"os ratio of $\eta=0.6(12)\times 10^{-5}$. This moderate value, compared with the current state of the art using two atomic species \cite{asenbaum_atom-interferometric_2020}, is mainly due to the gas temperature, which limits the free-evolution time to a few milliseconds. Longer free-evolution time could be achieved using for example delta-kicked cooling \cite{ammann1997delta}, and larger momentum kick thanks to shorter magic wavelength \cite{ma2025magic}. 

By adding a light-field gradient on the atom's path, we introduced a state-dependent interferometric method for differential polarizability measurements. We found a tune-out wavelength for the excited $^3P_0$ state at $478.95(8)$\,nm. Tune-out wavelengths are useful to extend the ultracold-gas quantum-simulation toolbox where state-dependent transport properties are required, in particular for Kondo lattice problems \cite{zhang2016kondo,nishida20133,bauer2013realizing,lange2025connecting}. 

In the future, magic wavelength dual matter-wave interferometers might play a crucial role as atomic clock interferometer to test post-Newtonian theories \cite{takamoto_test_2020,dimopoulos_testing_2007,roura_gravitational_2020,ufrecht_atom-interferometric_2020,di_pumpo_gravitational_2021}, and in searches for light bosons candidates of dark matter \cite{Arvanitaki_dark_matter_2018,Graham_DM_2016}. However, several refinements need to be made. (i) The sensitivity shall be improved, motivating the realization either of large-size terrestrial interferometers to increase the interrogation time \cite{dimopoulos_testing_2007,Abe_2021,Badurina2019_AION,El-Neaj2020_AEDGE}, or space-mission projects \cite{altschul_quantum_2015,tino_sage_2019}. (ii) Other types of matter-wave interferometers shall be considered since the MZI is insensitive to time dilation \cite{loriani_interference_2019,di_pumpo_gravitational_2021}. One possible approach consists of replacing the MZI with a modified Ramsey-Bord\'e interferometer, where atoms are spatially frozen in the middle of the interferometer sequence in an optical lattice for a certain time to allow phase accumulation in the gravitational potential~\cite{xu_probing_2019}.
(iii) the clock shall be activated either before \cite{roura_gravitational_2020} or during the interferometric sequence \cite{di_pumpo_gravitational_2021}. In both cases, the photon's recoil and differential light shifts shall be properly canceled to avoid uncontrolled phase shifts. 

Finally, we note that the weak equivalence test was performed using a clock-state statistical mixture. Using pure states instead, one can envision a genuine quantum test of the weak equivalence principle \cite{zych_quantum_2018} in the optical domain, namely several orders of magnitude larger than in radio-frequency domain where such a test has been previously reported \cite{rosi_quantum_2017}.

\section{Methods}\label{sec11}

\bmhead{Atomic sample preparation}
We prepared an ultracold sample of $^{88}\text{Sr}$ atoms using a two-stage magneto-optical trap (MOT), obtaining $1.0(1)\times10^7$ atoms at $1.0 (1)\ \mu\text{K}$. Then, the atoms were transferred into a 3D optical lattice operating at the $813\text{ nm}$-magic wavelength. 
After preparing a mixed state with approximately equal populations of the ground and excited clock states, we performed an adiabatic lattice ramp-down, which further cooled the remaining atoms to a temperature of $0.32(2)\,\mu$K. Additional details on the atomic sample preparation can be found in the Supplementary Information. Finally, we employed Bragg diffraction as a velocity-selective filter to reduce the vertical temperature to $0.15(2)\,\mu\text{K}$, while the transverse temperatures remain unchanged. The final atom number after velocity selection is $1.0(1)\times10^{5}$ while the momentum rms-value along the vertical interferometric axis is $0.67(4)\hbar k$. This value is smaller than the momentum transfer of $2\hbar k$ of the Bragg pulse, allowing for resolved Bragg peaks after TOF expansion (see insets in \fig{fig2}a).

\bmhead{Optical lattice and Bragg pulses}
The optical lattice is composed of two retro-reflected beams in the horizontal plane. The vertical optical lattice is formed by two co-propagating beams with orthogonal linear polarizations as shown in \fig{fig1}a. After passing through the cold atomic cloud, the beams are reflected by a concave mirror. A quarter-wave plate (QWP) in front of this mirror swaps the polarizations of the two beams upon retro-reflection and passing through the QWP a second time. The frequency difference $\Delta\omega$ between the two beams can be continuously tuned thanks to the acousto-optic modulator AOM$_2$, enabling the formation of movable lattice in the laboratory frame. For lattice operation, we need a static lattice with $\Delta\omega=0$. Then the relative phase of the two beams is locked such that their combined polarization is aligned with one principal axis of the QWP, leading to a standing wave with full contrast (see Supplementary Information for more details). For Bragg operation, a non-zero frequency difference is applied. Hence, two vertical standing waves move in opposite directions at a speed of $\Delta\omega/k_{\textrm{eff}}$ in the laboratory frame. 

\bmhead{Bragg atom interferometry}
The atom interferometer is implemented in a Mach-Zehnder configuration ($\pi/2 - \pi - \pi/2$) utilizing first-order Bragg diffraction. We applied a linear frequency chirp to one of the Bragg beams to compensate the phase shift due to Earth's gravitational acceleration, while maintaining a constant interrogation time $T$. 
The pulse sequence begins with a $25\ \mu\text{s}$ long $\pi/2$ pulse, which acts as a beam splitter. This pulse drives a transition between the $6.5\hbar k$ and $8.5\hbar k$ momentum states, effectively transferring $2\hbar k$ of momentum to half of the atomic population. The Bragg laser operates at the magic wavelength, ensuring that the Rabi frequency remains identical for both the ground $^1S_0$ and excited $^3P_0$ clock states. Following the initial split, the wavepackets propagate freely along two distinct trajectories for a free-evolution time $T$. A subsequent $50\ \mu\text{s}$ long $\pi$ pulse acts as a mirror pulse to exchange the momentum states of the wavepackets in the two interferometer arms. After a second interval $T$, a final $\pi/2$ pulse recombines the wavepackets. 

Population readout was performed using state-selective fluorescence imaging. After a TOF of $16\text{ ms}$, we first probed the $^1S_0$ population using a $30\ \mu\text{s}$ pulse of $461\text{ nm}$ light. To measure the $^3P_0$ population, we applied repumping lasers at $679\text{ nm}$ and $707\text{ nm}$ for $2\text{ ms}$ to return the atoms to the ground state. These atoms were then imaged using the $461\text{ nm}$ transition after a total TOF of $20\text{ ms}$. Each measurement is averaged over ten experimental cycles.

For the ground or the excited state, the TOF image shows two-peak distributions that correspond to the output of the MZI (see insets in \fig{fig2}a). The relative populations are extracted using two-dimensional Gaussian fits.

\backmatter

\bmhead{Acknowledgements}
This project is supported by the National Research Foundation, Singapore, through the National Quantum Office, hosted in A*STAR, under its Centre for Quantum Technologies Funding Initiative (S24Q2d0009), and the Singapore Ministry of Education Academic Research Fund Tier2 (Grant No. MOE-T2EP50223-0004).

\bmhead{Author contributions}
J. L., S. D. and X. M. developed and calibrated the experimental apparatus, performed the experiment, and collected and analysed the data. S. D., T. Z.-W., S.-Y. L., C. C. K. and D. W. interpreted the data. X. M., C. C. K. and D. W. developed the model to explain the data. All authors contributed to the writing of the manuscript. D. W. supervised the work.

\bmhead{Competing interests}
The authors declare no competing interests.

\bmhead{Data availability}
The data that support the findings of this study are available from the corresponding author upon reasonable request.

\putbib[Paper_Dual_Interferometer]
\end{bibunit}

\clearpage
\ResetForSupplementaryTitle
\setcounter{page}{1}
\setcounter{section}{0}
\setcounter{subsection}{0}
\setcounter{subsubsection}{0}
\setcounter{figure}{0}
\setcounter{table}{0}
\setcounter{equation}{0}
\renewcommand{\figurename}{Supplementary Fig.}
\renewcommand{\theequation}{S\arabic{equation}}
\renewcommand{\fig}[1]{Supplementary Fig.~\ref{fig:#1}}
\renewcommand{\refname}{Supplementary References}
\makeatletter
\providecommand{\theHpage}{}
\providecommand{\theHsection}{}
\providecommand{\theHsubsection}{}
\providecommand{\theHfigure}{}
\providecommand{\theHtable}{}
\providecommand{\theHequation}{}
\renewcommand{\theHpage}{SI.\arabic{page}}
\renewcommand{\theHsection}{SI.\arabic{section}}
\renewcommand{\theHsubsection}{SI.\arabic{section}.\arabic{subsection}}
\renewcommand{\theHfigure}{SI.\arabic{figure}}
\renewcommand{\theHtable}{SI.\arabic{table}}
\renewcommand{\theHequation}{SI.\arabic{equation}}
\makeatother

\begin{bibunit}[sn-mathphys-num]
\title[Article Title]{Supplementary Information for ``Magic-wavelength matter-wave interferometry with optical clock states"}

\author[1]{\fnm{Jianing} \sur{Li}}

\author[1,2,3]{\fnm{Swarup} \sur{Das}}

\author[3]{\fnm{Xinyuan} \sur{Ma}}

\author[4]{\fnm{Thomas} \sur{Zanon-Willette}}

\author[1,5,6]{\fnm{Shau-Yu} \sur{Lan}}

\author*[1,2,3]{\fnm{Chang Chi} \sur{Kwong}}\email{changchikwong@ntu.edu.sg}

\author[1,2,3]{\fnm{David} \sur{Wilkowski}}

\affil[1]{\orgdiv{Nanyang Quantum Hub, School of Physical and Mathematical Sciences}, \orgname{Nanyang Technological University}, \orgaddress{\street{21 Nanyang Link}, \city{Singapore}, \postcode{637371}, \country{Singapore}}}

\affil[2]{\orgdiv{MajuLab, International Joint Research Unit IRL 3654}, \orgdiv{CNRS, Universit\'e C\^ote d'Azur, Sorbonne Universit\'e, National University of Singapore, Nanyang Technological University} \country{Singapore}}

\affil[3]{\orgdiv{Center for Quantum Technologies}, \orgname{Nanyang Technological University Singapore}, \orgaddress{\street{50 Nanyang Avenue}, \city{Singapore}, \postcode{639798}, \country{Singapore}}}

\affil[4]{\orgname{Sorbonne Universit\'e CNRS, MONARIS, UMR 8233}, \orgaddress{\postcode{F-75005}, \state{Paris}, \country{France}}}

\affil[5]{\orgname{Department of Physics, National Taiwan University}, \orgaddress{\state{Taipei}, \country{Taiwan}}}

\affil[6]{\orgname{Institute for Atomic and Molecular Sciences, Academia Sinica}, \orgaddress{\state{Taipei}, \country{Taiwan}}}

\maketitle 

\section{Ultracold sample preparation and clock transition}

We transferred $1.0(1)\times10^7$ $^{88}$Sr atoms at a temperature of $1.0(1)\,\mu$K into a three-dimensional (3D) optical lattice operating at 813 nm magic wavelength. To generate the optical lattice, a $4.8$~W laser (Precilaser system) is split into three mutually orthogonal, retro-reflected beams that overlap at the atomic cloud. Each lattice arm has a fiber-coupled output power of 650(5) mW and a beam waist of 75(3) $\mu$m at the atoms. The mean trapping frequency is measured to be $\omega/(2\pi)=55(1)~$kHz, while the trap depth is $11.0(5)\,\mu$K. To ensure that the lattice beams do not interfere with each other, a frequency difference in the MHz range, namely much larger than the trapping frequency, is kept among the beams along the three directions. The atoms are confined in the Lamb-Dicke regime, with a Lamb-Dicke parameter of $\eta=\sqrt{w_r/w}=0.29(1)$, where $w_r/(2\pi)=4.64\,$kHz is the recoil frequency of the interrogating clock laser at 698 nm. We removed the weakly trapped atoms in the tails of the horizontal lattices by adiabatically ramping the lattice power to half in $10~$ms and then held them for $10~$ms before ramping the intensity back to the maximum in another $10~$ms. We ended up with about $1.0(1)\times 10^6$ atoms within a horizontal cloud size of $\sigma_{x,y} = 60(4)\,\mu$m and a vertical size of $\sigma_z = 35(2)\,\mu$m.

\begin{figure}
    \centering  \includegraphics[width=\linewidth]{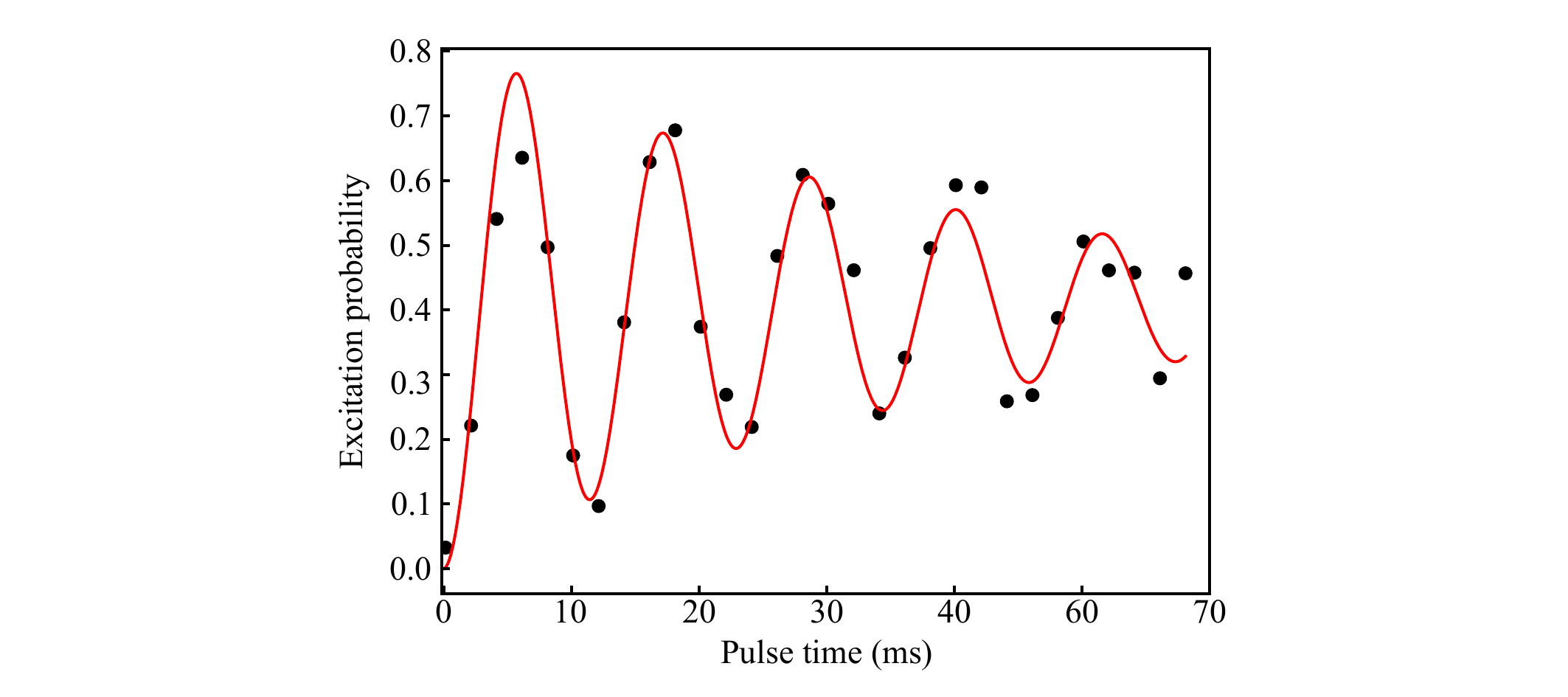}
    \caption{Observation of Rabi oscillations driving clock transition. Following the method in Ref.~\cite{Blatt_Rabi_2009}, the data are fitted with the function 
 $a \left[ 1 - \cos(2\pi f_{\text{Rabi}} t) \exp\left(-\frac{t}{\tau_{\text{c}}}\right) \right]$
 where $f_{\text{Rabi}} \sim 87.1(5)$ Hz represents the extracted atomic Rabi frequency and $\tau_{\text{c}}\sim 38(5)$ ms denotes the decoherence time scale.}
    \label{fig:SM_Fig1}
\end{figure}

A clock laser beam at 698 nm, derived from an external-cavity diode laser that is stabilized on an ultralow expansion cavity (Menlo system), is overlapped with one of the horizontal optical lattice beams using a dichroic mirror. The linewidth of the clock laser beam is expected to be below $10~$Hz.  Unlike the fermionic isotope, $^{88}$Sr has a zero nuclear spin; we applied a static magnetic field of $6~$mT along the polarization axis of the clock laser to activate the clock transition through a state mixing between the $5s5p~^3P_0$ and $5s5p~ ^3P_1$ states. The clock laser beam has a $360~\mu m$ waist and a $40~ $mW power, leading to a Rabi frequency slightly below $100~$Hz as shown in \fig{SM_Fig1}.
We prepared a mixed state with almost equal populations in both the ground and excited states. 

At the end of the clock pulse, the bias magnetic field is ramped down to zero, and the intensity of the lattices in all three dimensions are adiabatically switched off to further cool the atoms to $0.32(2)~\mu$K.
\section{Vertical lattice and Bragg laser}
\begin{figure}
    \centering  \includegraphics[width=\linewidth]{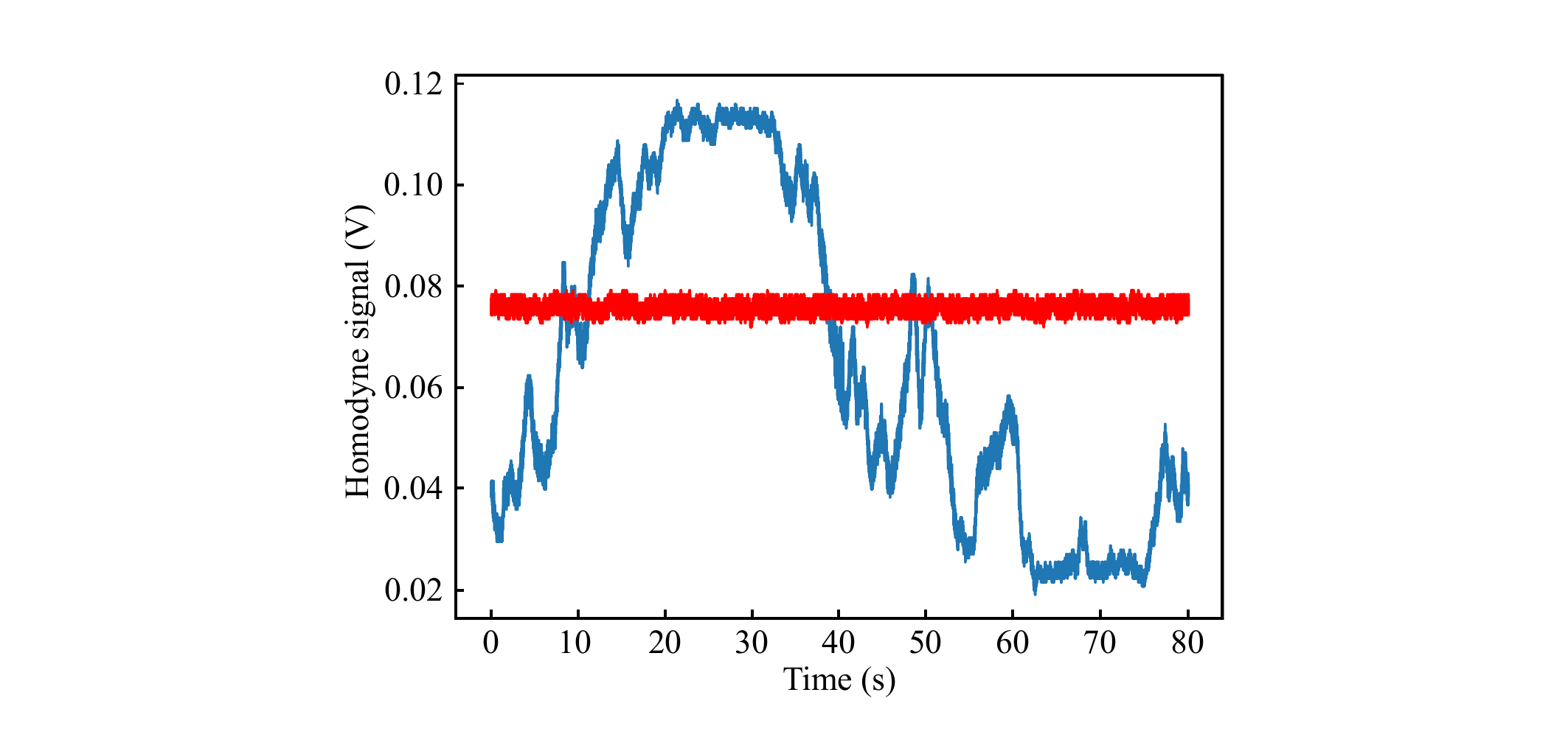}
    \caption{Relative phase variation of the two vertical lattice beams evaluated from the homodyne signal on the photo diode. The blue and the red traces show the phase fluctuations without and with the phase lock, respectively.}
    \label{fig:phase_lock}
\end{figure}

The vertical lattice consists of two beams with cross-polarizations focused on the atomic cloud and retro-reflected by a concave mirror. These beams serve as the lattice and the Bragg pulse. For that purpose, a quarter-wave plate is mounted before the retro-reflecting concave mirror to flip the polarization states of each beam (see Fig. 1a in the main text). On the path of each beam, we placed an acousto-optic modulator (AOM). On one of the beams, the AOM operates at a fixed $80~$MHz frequency. On the other beam, the AOM can be driven either by a voltage-controlled oscillator (VCO) for lattice operation, or by a direct digital synthesizer (DDS) during the Bragg pulse sequence.
During the lattice operation, we locked the two beams in phase to obtain a stable and high-contrast stationary wave. After passing through the AOMs, the two cross-polarized lattice beams are combined using a polarizing beam splitter (PBS) and then coupled into a polarization-maintenance fiber, which eventually degrades the relative phase stability. To recover optimum phase stability at the atomic cloud, a beam sampler is placed after the fiber output to extract a fraction of the combined beam for phase locking. Here, the beam is sent through a polarizer at 45 degrees and into a photodiode, see Fig. 1a in the main text. The homodyne signal is sent to a PID module to perform a feedback through the VCO. The electronic control loop is adjusted to keep a $\pi/2$ phase difference between the beams, namely where the derivative is maximum for an optimum locking performance. The latter is evaluated by examining the fluctuation of the photodiode signal, as illustrated in \fig{phase_lock}. Without a lock, the relative phase between the beams fluctuates over more than $2\pi$ within tens of seconds, whereas the fluctuation is suppressed below  $0.02\pi$ when the lock is engaged. 
To generate the Bragg pulses, a radio-frequency switch is used to switch from the phase-locked VCO output to the DDS output whose frequency and phase can be arbitrarily changed.

\section{Data analysis}

\begin{figure}
    \centering
    \includegraphics[width=\linewidth]{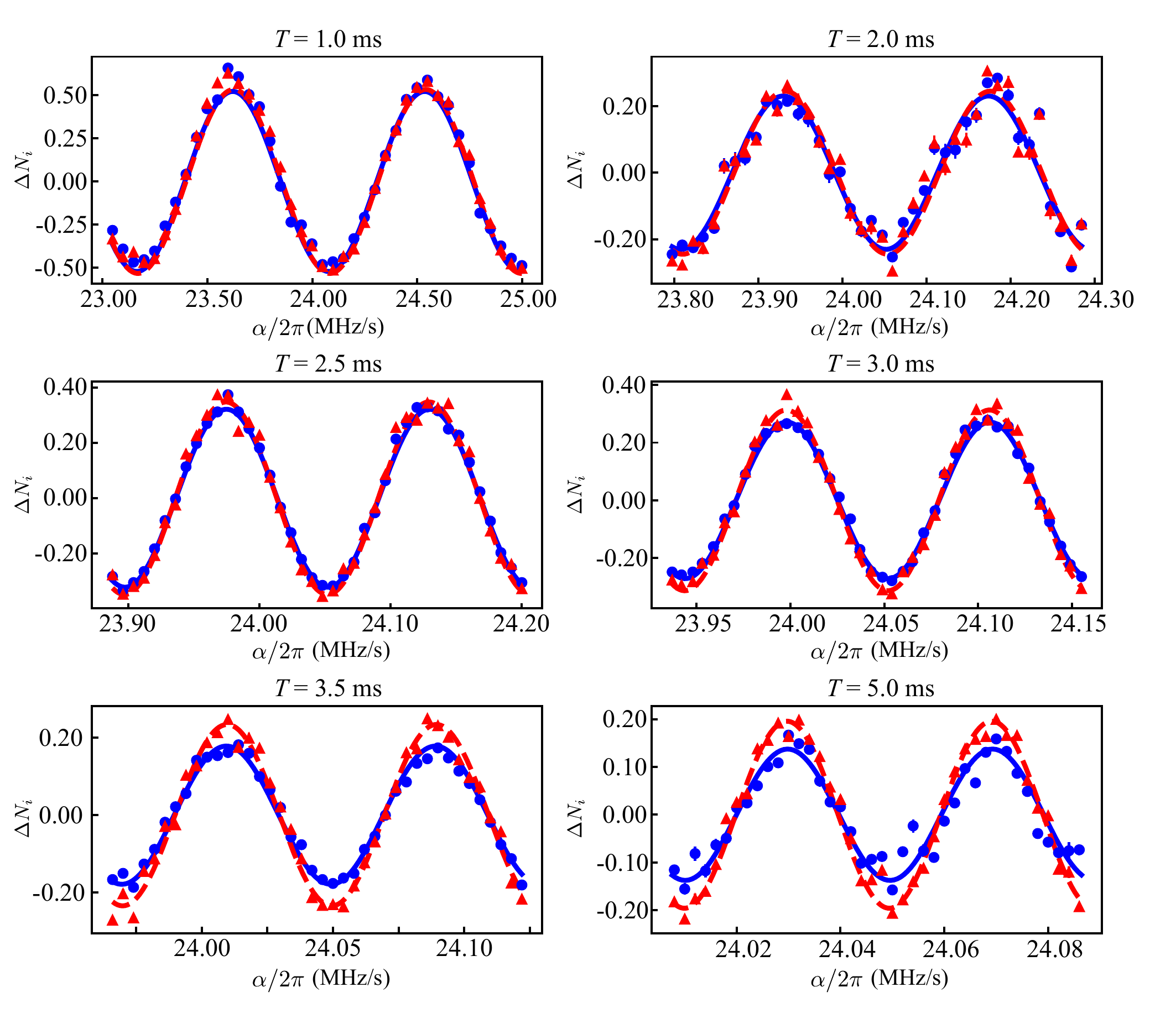}
    \caption{Dual interferometer signal obtained for both the ground (blue) and excited states (red) for several free-evolution times $T$ as function of chirp rate of the Bragg laser, each data point is averaged over ten experimental runs. The blue dot and red triangles represent ground and excited state interferometer signal and the solid and dashed lines represent a fit using \eq{SI_fitOutput}.}
    \label{fig:dual_AI}
\end{figure}

The interferometer fringes for each internal state are shown in \fig{dual_AI}. To evaluate the performance of the dual-state atom interferometer, we extracted two primary observables as a function of the interrogation free-evolution time $T$: the differential phase shift $\Delta\phi$ between the two internal state MZIs and their fringe visibilities. Fringes for each $T$ and each internal state are fitted independently with the function \begin{equation}
\Delta N_{i}= -V_i\cos{[(k_{\textrm{eff}}g-\alpha)T^2+\phi_i]},
\label{eq:SI_fitOutput}
\end{equation}   to extract their visibilities and phases. The subscript $i=g,e$ refers to the $^1S_0$ and $^3P_0$ states, respectively. Prior to fitting, the raw population difference is mean-centered to remove any signal  offset arising from atomic number imbalance between the two output ports. 

\begin{figure}
    \includegraphics[width=\linewidth]{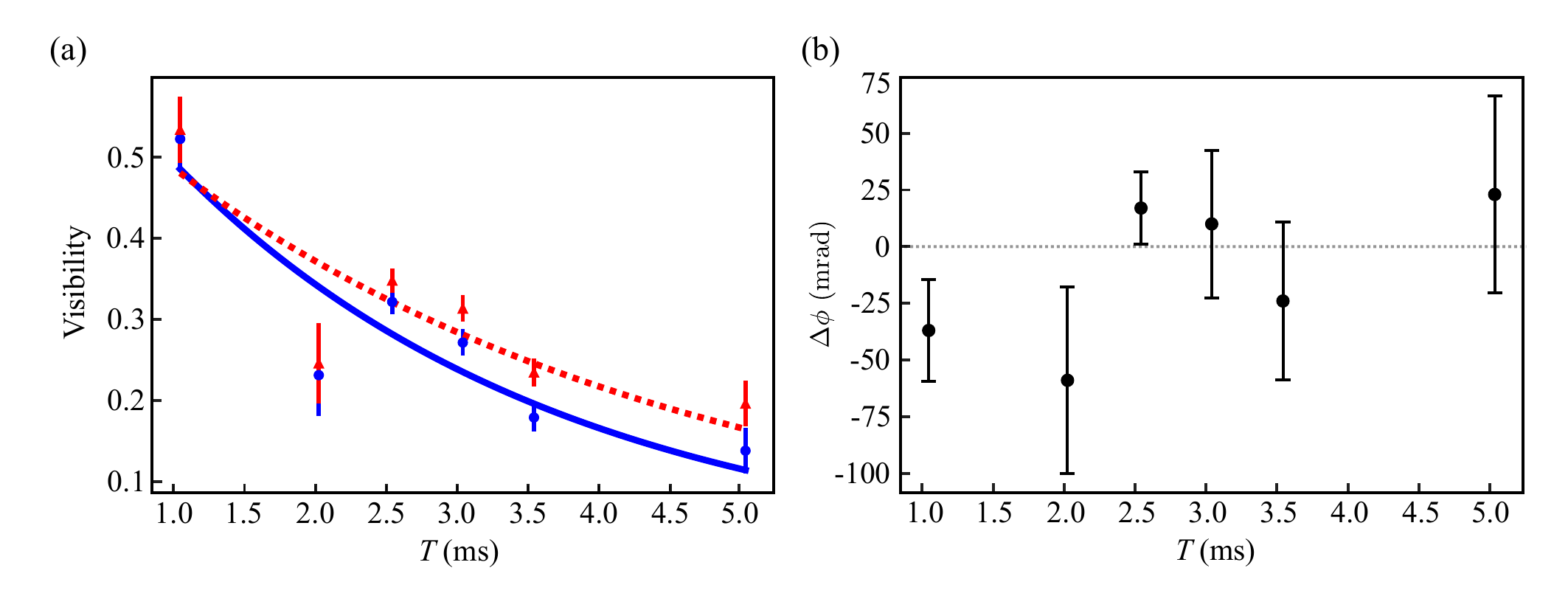}
    \label{fig:visibility}

 \caption{(a) Interferometer visibility of the ground state (blue dots) and excited state (red triangles) as function of free-evolution times. The data are extracted from sinusoidal fits to the MZI signal, shown in \fig{dual_AI}. The solid and dashed lines are exponential fits. (b) Phase difference vs free-evolution times. The data are extracted from the sinusoidal fits. The statistical errors are evaluated using a bootstrap method. }
\label{fig:visibility_phase}
\end{figure}

The free-evolution time $T$ is considered fixed from the electronically set value, leaving the visibility and phase as free parameters. The latter are determined by a weighted nonlinear least-squares-fitting method. The visibilities of the interferometer for different free-evolution times are fitted with an exponential function which determines the visibility decay time constants for the ground and the clock states to be $2.7(7)\,$ms and $3.7(1)\,$ms, respectively (see \fig{visibility_phase}a). This rapid decay is likely due to the finite temperature of the atomic cloud: during the free-evolution time, thermal expansion causes the cloud to grow to a size comparable to that of the interferometer beams. As a result, the atoms sample wavefront aberrations and intensity inhomogeneities across the beam profile. 

From the fitted phase values, the phase shift difference between the two MZIs was computed from $\Delta\phi = \phi_{g}-\phi_{e}$, and mapped onto the $[-\pi, \pi]$ range. These values are plotted against $T$ in \fig{visibility_phase}b.

To estimate the statistical uncertainties of $\Delta \phi$ that are plotted as the error bars, we employed a residual bootstrap procedure~\cite{Bootstrap_1}. For each free-evolution time, the best-fit curve was subtracted from the data to obtain the fit residuals. Over $M = 2000$ iterations, these residuals are resampled with replacement and added back onto the original fitted curve at the fixed design points, generating synthetic fringe datasets. The residuals are centered before resampling. Each synthetic dataset was then independently refitted with the \eq{SI_fitOutput} for both internal states and the replicate differential phase $\Delta \phi_{sub,m}$ was computed for each iteration. The reported experimental uncertainty $\sigma_{\Delta \phi}$ is the standard deviation of the resulting bootstrap ensemble:
$$\sigma_{\Delta\phi} = \sqrt{\frac{1}{M-1} \sum_{m=1}^{M} \left( \Delta\phi_{\text{sub}, m} - \langle\Delta\phi_{\text{sub}}\rangle \right)^2}.$$ 

From this set of phase difference values,  we perform a weighted average of the phase difference to find $|\overline{\Delta\phi}|=12(12)$~mrad as reported in the main text for our dual MZI. For an experiment performed at a given $T$, the typically phase uncertainty is given by 30~mrad, setting the sensitivity of our dual MZI.

The ultimate phase sensitivity of the dual interferometer is fundamentally constrained by statistically independent quantum projection noise. For a total of $n=400$ measurements (10 experimental runs each with 40 chirp rate values), the differential shot-noise limit after averaging over the 10 sets is expressed by $ \frac{1}{\sqrt{n}} \sqrt{\frac{1}{V_g^2 N_g} + \frac{1}{V_e^2 N_e}} \approx 2 \text{ mrad}$, with $N_{g,e}$ and $V_{g,e}$ representing the respective atom populations and fringe visibility of the ground and excited states at the $T= 5\text{ ms}$. However, this fundamental limit remains lower than our actual measured phase error of $\sim30\text{ mrad}$, which is currently dominated by technical noise sources.
Sensitivity can be pushed closer to the shot-noise limit by improving the cooling of the atomic cloud and increasing the momentum separation. This greater separation can be achieved using techniques like a shorter magic wavelength~\cite{SI_ma2025magic} and optimized pulse shaping based on optical control theory.

\section{Tune-out frequency measurement}
\begin{figure}
    \centering
    \includegraphics[width=\linewidth]{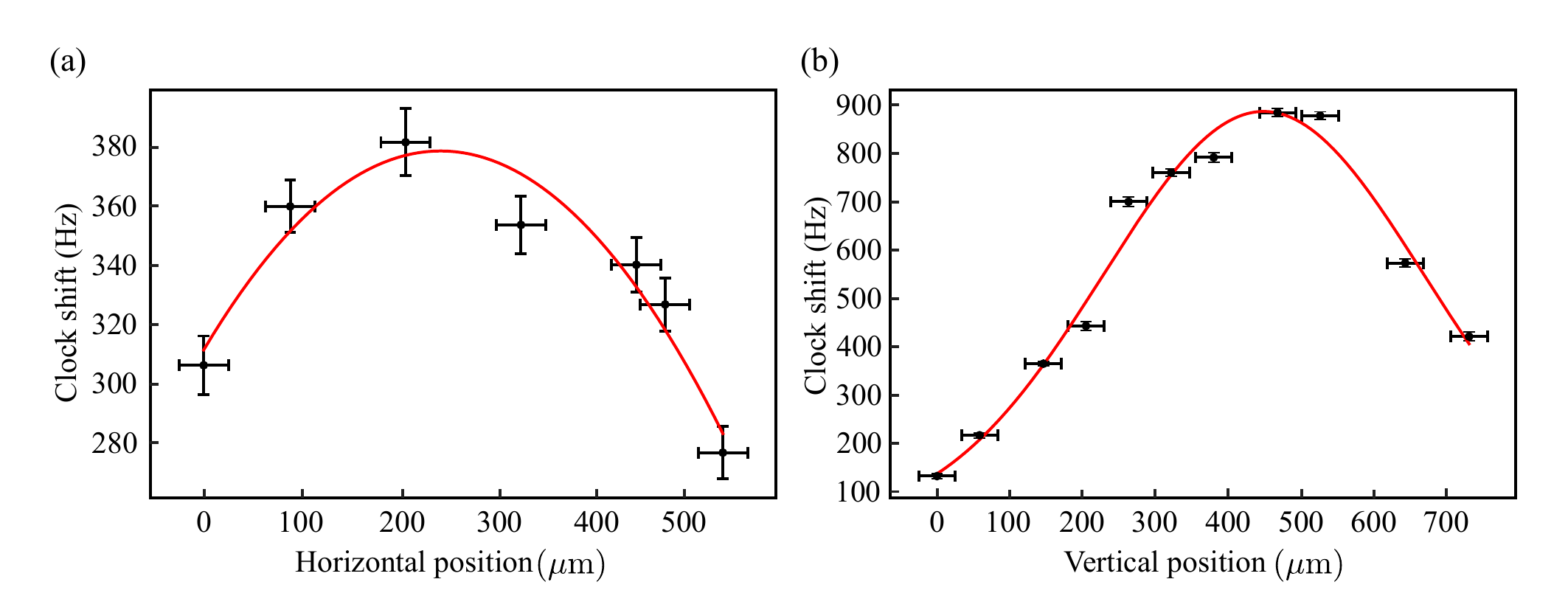}
    \caption{Spatial characterization and alignment of the $479\text{~nm}$ laser beam. (a) Shift of the clock transition frequency as a function of the horizontal beam position, fitted with a Gaussian profile to locate the horizontal center. (b) Vertical position scan mapping out the spatial profile. The beam position is subsequently positioned near $w_0/2$ relative to the atomic cloud center to maximize the local optical gradient along the interferometer direction. }
    \label{fig:position_479}
\end{figure}
The dual-state MZI, running simultaneously for two internal states, facilitates the probing of state-dependent forces through the measurement of their differential phase. We illustrated this capability by measuring the tune-out wavelength of the excited clock state $^3P_0$ near $479~\text{nm}$.

To apply a state-dependent force, a laser beam with a waist of $w_0 = 620(5)~\mu\text{m}$ is turned on during the interferometer's free-evolution time. This beam induces a dipole potential $U_i(\textbf{r}) = -\frac{\alpha_i(\omega) I(\textbf{r})}{2 \epsilon_0 c}$ for an atom in state $i \in \{^1S_0,\ ^3P_0\}$, where $\alpha_i(\omega)$ is the polarizability at the optical frequency $\omega$, and $I(\textbf{r}) = I_0 e^{-2r^2/w_0^2}$ is the Gaussian intensity profile with peak intensity $I_0$. This potential is in the transverse plane of the beam and results in a transverse acceleration:
\begin{equation}
   \textbf{a}_{i}(\textbf{r}) = -\frac{\nabla U_i(\textbf{r})}{m} = -\frac{2 \textbf{r}}{\epsilon_0 c w_0^2 m} \alpha_{i}(\omega) I_0 e^{-2r^2/w_0^2},
   \label{eq:gaussianAcc}
\end{equation}
which adds to the Earth's gravitational acceleration for a horizontal beam.

Due to the AC Stark shift, the resonance center of the clock transition shifts in the presence of the $479~\text{nm}$ beam. This shift in the clock transition frequency is utilized to align the beam with the atomic cloud. Following an initial rough alignment, we performed a calibrated horizontal position scan in steps of $100~\mu\text{m}$ to locate the position of the maximum shift. Keeping the horizontal position fixed at this maximum, we scanned the vertical position as shown in \fig{position_479}. Finally, to maximize the optical gradient force, we positioned the beam vertically such that the atomic cloud is close to the $w_0/2$ point where the intensity gradient is the highest.

With the beam aligned, we measured the phase accumulated by the excited state as a function of the $479~\text{nm}$ laser wavelength. The phase is compared to the bare interferometer phase (without the 479~nm beam) for a fixed free-evolution time of $T = 5~\text{ms}$. For each wavelength, the measurement was repeated 5 times to obtain the average phase shift. The tune-out wavelength is extracted by performing a linear fit on the experimental data (see Fig.~3 in the main text). Finally, we compared our experimental data to a theoretical model, which is described in the next section.

\section{Estimation of MZI phase difference with optical gradient force}

In the tune-out wavelength measurement experiment, the atoms move over a distance of around 160~$\mu$m during the full MZI sequence. This distance is not small compared to the 620~$\mu$m of the 479~nm laser beam waist, leading to non-uniform acceleration $a(z)$ [see \eq{gaussianAcc}]. In this section, we detail the derivation of the interferometric phase in this context. 

We employed the perturbative approach of Ref.~\cite{storey_feynman_1994} to perform our calculations. In this calculation, we are concerned only with the vertical degree of freedom. The Lagrangian of the system is given by $L=L_0+L_p$, where
\begin{align}
    L_0 =& \frac{1}{2}m\dot{z}^2-mgz,\\
    L_p =& -U(\mathbf{r}).
\end{align} 
$z$ denotes the vertical position of the atom, and $\dot{z}$ is the velocity component of the atom along $z$. The term $L_0$ is the unperturbed Lagrangian of a particle undergoing free-fall in gravity. The term $L_p$ is the perturbation in the Lagrangian arising from the optical force. The MZI phase difference between the two arms is given by $\Delta\Phi^{p}_{total} = -k_{\textrm{eff}}gT^2 + \Delta\Phi_{pert}$,
where
\begin{equation}
    \Delta\Phi_{pert}= \frac{1}{\hbar}\int_{t_i}^{t_f} [L_P(z_{up}(t)) - L_P(z_{low}(t))]~dt.
\end{equation}
The initial and final times of the MZI sequence are denoted as $t_i$ and $t_f$, respectively. The integration is performed over the unperturbed classical trajectories of the upper and the lower arms of the interferometer, which are denoted by $z_{up}$ and $z_{low}$, respectively. Here, these trajectories are simply the free-fall trajectories of the atoms, taking into account the initial positions and velocities of the atoms.

\begin{figure}
    \centering\includegraphics[width=\linewidth]{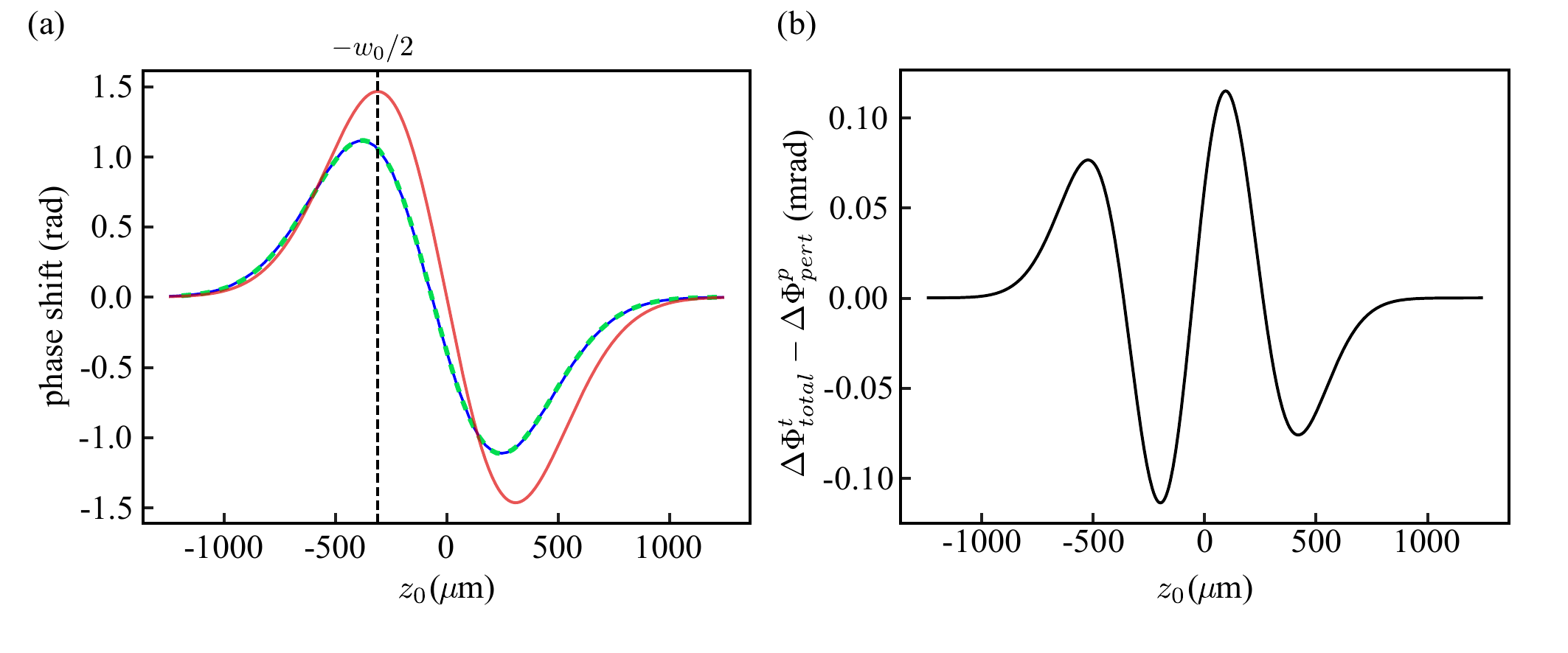}
    \caption{(a) Contribution of the optical force to the MZI phase at $T=5$~ms for $^3P_0$ state, plotted as a function of the initial vertical position.  The red solid line shows the phase assuming that the atoms experience a uniform acceleration of $a(z_0)$ during the MZI sequence, while the blue solid line accounts for the position-dependent force along the atomic trajectory. The green dashed line includes the effects of the finite cloud size and thermal velocity spread. (b) The difference between the total MZI phase calculated using the perturbative method, $\Delta\Phi^p_{total}$, and the phase calculated using the actual classical trajectory, $\Delta\Phi^t_{total}$. We considered the single-atom case here, with the atom initially at rest. 
    In both panels, we set laser power $P_0 = 48~ \text{mW}$, the waist $w_0=620~\mu\text{m}$ and the laser wavelength $\lambda = 478.00$ nm.}
    \label{fig:phase_478}
\end{figure}

Using this formalism, we computed $\Delta\Phi_{pert}$ as a function of $z_0$, the initial vertical position of the atom, and plotted it as the blue solid line in \fig{phase_478}(a). Here, we took a free-evolution time of $T=5$~ms to compute the phase shift experienced by the $^3P_0$ state. A comparison was made with the case where the acceleration due to the optical force is assumed to be constant at $a(z_0)$ throughout the MZI sequence (red solid line). Accounting for the non-uniform acceleration leads to a phase difference of up to 0.5 rad. Due to the motion of the atoms during the MZI sequence, the initial position of the atoms to achieve the maximum sensitivity to the optical force is shifted slightly from $z=\pm w_0/2$. Since we placed the atoms at a position of $-w_0/2$, there is only a 5\% reduction in the phase shift due to the optical force, which does not significantly affect our experiment.

Instead of using the perturbative calculations outlined above, we can also directly solve the following equation of motion for the actual classical trajectories of the atoms in the presence of the optical force. 
\begin{equation}
\label{eq:EOM}
m\ddot{z} = -mg - \frac{2z}{\epsilon_0 cw_0^2}\alpha_{i}(\omega)I_0 e^{-2z^2/w_0^2}
\end{equation}
Using the classical trajectories, we can compute the MZI phase using the following decomposition~\cite{hogan2009lightpulse,storey_feynman_1994},
\begin{equation}
\Delta\Phi^{t}_{total}=\Delta\Phi_{propagation}+\Delta\Phi_{laser}+\Delta\Phi_{separation}.
\end{equation}
The propagation phase is proportional to the difference in the action of the classical trajectories along the two arms. While this phase is usually zero for the case of free fall, here it is non-zero. The non-uniform force also results in a separation of the two arms at the end of the MZI sequence, resulting in non-zero separation phase. Finally, as in the usual case, the interaction of atoms with the laser pulses also introduces a phase shift to the component of the state that undergoes a change in momentum. By summing all three contributions, we calculated $\Delta\Phi^{t}_{total}$ as a function of $z_0$, using our experimental parameters. We computed the difference between this method and the perturbative method, finding them to be very small at $\sim 0.1$ mrad compared to our experimental sensitivity of $\sim30$ mrad, as shown in \fig{phase_478}(b). This validates the perturbative approach used in our calculations. 

Finally, we investigated the effects coming from the spatial and velocity distributions of the atomic ensemble. The finite temperature of the atomic cloud gives rise to a thermal velocity distribution, which we modeled using Boltzmann statistics. The spatial distribution of the cloud is described by a two-dimensional Gaussian with vertical and horizontal radii of $35~\mu\text{m}$ and $60~\mu\text{m}$, respectively. To account for both effects, we sampled 10,000 atoms from these position and velocity distributions and computed the corresponding ensemble-averaged phase. As shown by the green dashed line in \fig{phase_478}(a), these effects introduce only a small correction, below 1\%. This shows that our atomic cloud is sufficiently small and cold such that it does not significantly affect the MZI phase. 

Using the same method and taking the wavelength dependence of the $^3P_0$ state polarizability into account, we computed the MZI phase as a function of 479~nm laser frequency. This is plotted in Fig. 3b of the main text for comparison with experimental data.

\putbib[Paper_Dual_Interferometer]
\end{bibunit}

\end{document}